\documentclass[11pt]{article}
\usepackage[T1]{fontenc}
\usepackage{newpxtext}
\usepackage{comment}
\usepackage{mathpazo}
\usepackage{amsmath, amsthm, amsfonts, amssymb, graphicx}
\usepackage{dsfont}
\usepackage[a4paper]{geometry}
\usepackage[full]{complexity}
\usepackage{hyperref}
\usepackage{comment,bbm,enumitem}
\usepackage{xcolor}

\DeclareMathOperator{\mindeg}{{\mbox{\rm \textsf{mindeg}}}}
\DeclareMathOperator{\val}{{\mbox{\rm \textsf{val}}}}
\DeclareMathOperator{\mval}{{\mbox{\rm \textsf{minval}}}}

\DeclareMathOperator{\rank}{{\mbox{\rm rank}}}

\DeclareMathOperator{\diag}{{\mbox{\rm Diag}}}
\renewcommand{\epsilon}{\varepsilon}

\newcommand{\e}{\epsilon}
\newcommand{\vect}[1]{\mathbf{#1}}
\newcommand{\F}{\mathds{F}}
\newcommand{\N}{\mathds{N}}
\newcommand{\Z}{\mathds{Z}}
\renewcommand{\R}{\mathds{R}}

\newcommand{\infZ}{\mathds{Z}\cup\{+\infty\}}
\renewcommand{\C}{\mathds{C}}
\newcommand{\calB}{\mathcal{B}}

\newcommand{\I}{\mathcal{I}}

\newcommand{\VBP}{\mbox{\rm{\textsf{VBP}}}}
\newcommand{\VF}{\mbox{\rm{\textsf{VF}}}}
\renewcommand{\VNP}{\mbox{\rm{\textsf{VNP}}}}

\newcommand{\calI}{\mathcal{I}}

\newtheorem{theorem}{Theorem}[section]
\newtheorem{corollary}[theorem]{Corollary}
\newtheorem{proposition}[theorem]{Proposition}
\newtheorem{lemma}[theorem]{Lemma}

\newtheorem{claim}[theorem]{Claim}

\newtheorem{observation}[theorem]{Observation}
\newtheorem{remark}[theorem]{Remark}

\newcommand{\claimproof}[2]%
{\noindent{\em Proof of Claim \ref{#1}.}
#2\hspace*{\fill}$\Box$~~~~~\vspace{5mm} }

\title{\textbf{Border Complexity of Symbolic Determinant under Rank One Restriction}}
\author{
Abhranil Chatterjee\thanks{ACM Unit, Indian Statistical Institute, Kolkata, Email: \texttt{abhneil@gmail.com}} \and Sumanta Ghosh\thanks{Department of Computing and Mathematical Sciences, Caltech, Email: {\texttt besusumanta@gmail.com}}\and Rohit Gurjar\thanks{Department of Computer Science and Engineering, IIT Bombay, Email: \texttt{rgurjar@cse.iitb.ac.in}} \and Roshan Raj\thanks{Department of Computer Science and Engineering, IIT Bombay, Email: \texttt{roshanraj@cse.iitb.ac.in}}}
\date{}

\begin{document}
\maketitle  
\begin{abstract}
$\VBP$ is the class of polynomial families that can be computed by the determinant of a symbolic matrix of the form $A_0 + \sum_{i=1}^n A_ix_i$ where the size of each $A_i$ is polynomial in the number of variables (equivalently, computable by polynomial-sized algebraic branching programs (ABP)). A major open problem in geometric complexity theory (GCT) is to determine whether $\VBP$ is closed under \emph{approximation} i.e.\ whether $\VBP \stackrel{?}{=}\overline{\VBP}$. The power of approximation is well understood for some restricted models of computation, e.g.\ the class of depth-two circuits, read-once oblivious ABPs (ROABP), monotone ABPs, depth-three circuits of bounded top fan-in, and width-two ABPs. The former three classes are known to be closed under approximation~\cite{DBLP:conf/coco/BlaserIMPS20}, whereas the approximative closure of the last one captures the entire class of polynomial families computable by polynomial-sized formulas~\cite{BIZ17}. 

In this work, we consider the subclass of $\VBP$ computed by the determinant of a symbolic matrix of the form $A_0 + \sum_{i=1}^n A_ix_i$ where for each $1\leq i \leq n$, $A_i$ is of rank one. This class has been studied extensively~\cite{Edm68, Edm79, Mur93} and efficient identity testing algorithms are known for it~\cite{Lov89, GT20}. We show that this class is closed under approximation.
In the language of algebraic geometry,
we show that the set obtained by taking coordinatewise products of pairs of points from (the Pl\"{u}cker embedding of) a Grassmannian variety is closed.

\end{abstract}
%\chapter{A Deterministic Parallel Reduction from Weighted Matroid Intersection Search to Decision }
\newpage
\section{Introduction}

The determinant polynomial plays a central role in the study of complexity theory. It is known to be a \emph{complete polynomial} i.e. every polynomial can be computed by some affine projection of the determinant of a symbolic matrix.  More precisely, for any polynomial $f\in \F[x_1, \ldots, x_n]$, there is some $m$ and $A_0, A_1, \ldots, A_n$ in $\F^{m\times m}$ such that $f = \det_m(A_0 + \sum_{i=1}^n A_ix_i)$. $\VBP$ is defined as the class of polynomial families for which the size of such determinantal representation is polynomially bounded in the number of variables (equivalently, such polynomial families can be computed by polynomial-size \emph{algebraic branching programs} (ABP)).

The other polynomial of significant interest is the permanent polynomial, a close cousin of the determinant polynomial. The permanent polynomial is also known to be a \emph{complete polynomial}. $\VNP$ is defined as the class of polynomial families for which the size of the permanental representation is polynomially bounded in the number of variables. It is known that $\VBP \subseteq \VNP$. The goal of algebraic complexity theory is to separate $\VBP$ and $\VNP$, equivalently, to show a super-polynomial lower bound on the determinantal representation of the permanent polynomial.

Even though we have witnessed some outstanding progress in our understanding of the lower bound problem on various restricted models of computation in the last few years, the fundamental problem in the general setting remains elusive. \emph{Geometric Complexity Theory} (GCT) was proposed as a possible approach to settle this question by showing $\VNP \not\subseteq \overline{\VBP}$~\cite{MS01} where 
 $\overline{\VBP}$ denotes the \emph{approximative closure} of $\VBP$. Let $\mathcal{C}$ be a circuits class over $\F$, $\F[\epsilon]$ be the polynomial ring and $\F(\epsilon)$ be the fraction field of $\F[\epsilon]$. We can define $\overline{\mathcal{C}}$, the (approximative) closure of the circuit class $\mathcal{C}$ in the following equivalent ways.
 
\noindent\textit{(a) Approximative closure.} 
A polynomial family $\{f_n\}$ is in the approximative closure of $\mathcal{C}$ over $\F$ if there is a polynomial family $\{g_n\}$ in $\F[\epsilon][x_1, \ldots, x_n]$ computable in $\mathcal{C}$ over $\F(\epsilon)$, such that for every $n$, 
\[
g_n(x_1, \ldots, x_n) = f_n(x_1, \ldots, x_n) + \epsilon \cdot h_n(x_1, \ldots, x_n)
\]
for some polynomial $h_n$ in $\F[\epsilon][x_1, \ldots, x_n]$. We say, the polynomial family $\{f_n\}$ is approximated by the family $\{g_n\}$.

\noindent\textit{(b) Euclidean closure.}
A polynomial family $\{f_n\}$ is in the Euclidean closure of $\mathcal{C}$ over $\F$ if, for every $n$, there exists an infinite sequence of polynomials $\{g_{n,i}\}$ in $\mathcal{C}$ over $\F$ such that the limit point of the sequence of coefficient vectors corresponding to $\{g_{n,i}\}$ is the coefficient vector of $f_n$. This definition is known to be equivalent to the previous definition when $\F$ is $\mathbb{R}$ or $\mathbb{C}$~\cite{Bur04}. 

\noindent\textit{(c) Zariski closure.}
Another equivalent way is to define the approximative closure as a \emph{Zariski closure}~\cite{Mum95}. For a circuit class $\mathcal{C}$, consider the system of all polynomial equations 
%of the largest size 
%such that each equation is
which are satisfied by the coefficient vector corresponding to each polynomial in $\mathcal{C}$. Then, the Zariski closure $\overline{\mathcal{C}}$ consists of the polynomials such that the corresponding coefficient vectors are satisfying assignments of the system of polynomial equations. 

%A polynomial $P$ in $\F[x_1, \ldots, x_n]$ is approximately computed by $Q(x_1, \ldots, x_n, \epsilon)$ in  $\F[\epsilon][x_1, \ldots, x_n]$ if there exists $R(x_1, \ldots, x_n, \epsilon)$ in  $\F[\epsilon][x_1, \ldots, x_n]$ such that $Q = P + \epsilon R$. It can also be stated as $\lim_{\epsilon\to 0} Q = P$ where the limit point is defined using the Euclidean topology. Now given a circuit class $\mathcal{C}$, $\overline{\mathcal{C}}$ is defined as the topological closure i.e.\ a polynomial $P$ is in the approximative closure $\overline{\mathcal{C}}$, if there is a polynomial $Q$ in  $\F[\epsilon][x_1, \ldots, x_n]$ that can be computed by a small circuit in $\mathcal{C}$ over $\F(\epsilon)$ and $\lim_{\epsilon\to 0} Q = P$. %The $\overline{size}(P)$ is the minimum over the size of the circuits computing some $Q$ approximately computing $P$. 

As all these definitions are equivalent, without loss of generality, we define $\overline{\mathcal{C}}$ to be the approximative closure of $\mathcal{C}$. If $\mathcal{C} = \overline{\mathcal{C}}$, we say $\mathcal{C}$ is closed under approximation. 

One of the main objectives of geometric complexity theory is to decide whether $\VBP$ is closed under approximation or not. Showing $\VBP = \overline{\VBP}$ would imply that showing $\VBP\neq \VNP$ is equivalent to showing $\VNP\not\subseteq \overline{\VBP}$. 
% A refutation would also have interesting consequences. Almost all the known lower bound techniques are known to be oblivious to the approximative closure~\cite{Gro13}. $\VBP \neq \overline{\VBP}$  would imply that if an existing lower-bound technique can separate $\VBP$ and $\VNP$, then it would even separate $\overline{\VBP}$ from $\VNP$. 
Though the complexity of $\overline{\VBP}$ is not well-understood, the power of approximation has been successfully studied for various restricted models of computation. For example, it is known that the following classes are closed under approximation: (a) $\Sigma\Pi$ i.e. the sparse polynomials, (b) Monotone ABPs~\cite{DBLP:conf/coco/BlaserIMPS20}, and (c) Read-once oblivious ABPs (ROABP). Recently, the approximative closure of the depth three circuits of bounded top fan-in is shown to be contained in $\VBP$~\cite{DS22}. Surprisingly, even a restricted circuit class can efficiently compute a much larger class under approximation. For example, consider $\VBP_2$, the class of polynomials computed by the width-two ABPs. Even though, there are families of polynomials that cannot be expressed by this class ~\cite{AW16} , the approximative closure of this class contains $\VF$, the class of polynomials computed by a small formula. Indeed, it is known that $\overline{\VBP_2} = \overline{\VF}$~\cite{BIZ17}.

It is interesting to notice that, for the circuit classes for which the approximative closure is well-understood, we also know efficient identity testing algorithms. %for those classes. 
It motivates us to study the class VBP under some natural restriction for which we already have an efficient identity testing algorithm. The class of our interest is the symbolic determinant under rank one restriction. Recall that any $n$-variate polynomial in VBP can be computed as $\det(A_0 + \sum_{i=1}^n A_ix_i)$ where the size of each $A_i$ is polynomially bounded in $n$. We consider the class of polynomials of form $\det(A_0 + \sum_{i=1}^n A_ix_i)$ where for each $1\leq i \leq n$, $\rank(A_i) = 1$. This class has been studied extensively in contexts of polynomial identity testing, combinatorial optimization, and matrix completion (see, for example~\cite{Edm67, Lov89, Mur93}).  
It admits a deterministic polynomial-time identity testing algorithm in the white-box setting~\cite{Lov89} and a deterministic quasi-polynomial-time algorithm in the black-box setting~\cite{GT20}. This class is equivalent to the class of polynomial families computed by the determinant of symbolic matrices with each variable occurring at most once, also known as read-once determinants \cite{PC} (as cited in \cite[Lemma 4.3]{GT20}).
The expressive power of this class has also been studied. It strictly contains some well-studied classes like the polynomials computed by a small read-once formula (see, for example~\cite{DBLP:conf/fct/AravindJ15}).
%
%Some limits on the expressive power of this class have also been shown. 
%
%For example,
However, it is known that for large enough $n$, $n$-variate elementary symmetric polynomials 
and the permanent polynomial cannot be expressed 
as $\det(A_0+\sum_{i=1}^n A_ix_i)$
with $\rank(A_i)=1 \text{ for each } i\in [n]$ \cite{DBLP:conf/fct/AravindJ15}. %Formally, there doesn't exist $A_0,A_1,\dots,A_n$ such that $\det(A_0 + \sum_{i=1}^n A_ix_i)$ is equal to the permanent polynomial in $n$ variables.

Another motivation to study the approximative closure of this class 
%is far too ambitious. It is known
is the fact that the approximative closure of the orbit of this class under the action of the general linear group contains $\VBP$ \cite{MS21, ST21}. Therefore, understanding the approximative closure of this class may shed new light on the $\VBP\stackrel{?}{=} \overline{\VBP}$ question.

\paragraph*{Our Results.}
The main result of this paper is that the class of the determinant of symbolic matrices under rank one restriction is closed under approximation. More precisely, we show the following theorem, where we use $\F$ to denote $\R$ or $\C$. 

\begin{theorem}\label{theorem:main}
Given $A_0, A_1, A_2, \ldots, A_n \in {\F(\epsilon)^{r\times r}}$ such that for each $1\leq i\leq n$, $\rank(A_i) = 1$ over $\F(\epsilon)$. Let $f = \lim_{\epsilon\to 0}\det(A_0 + \sum_{i=1}^n A_ix_i)$ be defined. Then, there exists $B_0, B_1, B_2, \ldots, B_n$ in  ${\F^{(n+r)\times (n+r)}}$
such that $f = \det(B_0 + \sum_{i=1}^n B_ix_i)$ and $\rank(B_i) = 1$ over $\F$ for each $i\in [n]$. Moreover, if $A_0 = 0$, then	the matrices $B_1, B_2, \ldots, B_n$ lie in ${\F^{r\times r}}$. 
\end{theorem}

Since this class is closed under approximation, the known hitting set and non-expressibility results for this class also hold for its approximative closure. 

\begin{remark}
    By using formal power series, we can extend this result to any arbitrary field. For the sake of simplicity, we only work with $\C \text{ or } \R$.
\end{remark}
    
%\begin{theorem}\label{theorem:main}
%Given $A_1, A_2, \ldots, A_n \in {\F(\epsilon)^{r\times r}}$ such that $\rank(A_i) = 1$ over $\F(\epsilon)$ for each $i\in [n]$. Let $f = \lim_{\epsilon\to 0}\det(\sum_{i=1}^n A_ix_i)$ be defined. Then, there exists $B_1, B_2, \ldots, B_n \in {\F^{r\times r}}$ such that $f = \det(\sum_{i=1}^n B_ix_i)$ and $\rank(B_i) = 1$ over $\F$ for each $i\in [n]$.	
%\end{theorem}

\paragraph*{An algebraic geometry perspective on the result.}
Consider the simpler case of Theorem~\ref{theorem:main}, when $A_0=0$.
Using known techniques, the statement can be reduced to this simpler case. 
Now, suppose $A_1, A_2, \dots, A_n$ are $r \times r$ matrices of rank $1$. %
Let us write $A_i = \vect{u}^i \cdot {\vect{v}^i}^T$ for some vectors $\vect{u}^i, \vect{v}^i \in \F^{r}$ and define
matrices $U, V \in \F^{r \times n}$ whose
$i$th columns are $\vect{u}^i$ and $\vect{v}^i$, respectively. 
%
%The determinant $\det(\sum_i A_i x_i$
%can be equivalently written as $\det(UXV^T)$,
%where $X$ is the $n\times n$ diagonal matrix %with $x_1, x_2, \dots, x_n$ on the diagonal. 
%
It can be verified that  
\[\det\left(\sum_i A_ix_i\right) =
\sum_{S} \det(U_S) \det(V_S) \prod_{j \in S} x_j,\]
where the sum is over all 
size-$r$ subsets $S$ of $[n]$ and $U_S$ (or $V_S$) denotes the submatrix of $U$ (or $V$) obtained by taking columns with indices in the set $S$. 
Hence, essentially our main result says that the image of  the map 
\[(\F^{r \times n})^2 \to \F^{\binom{n}{r}}, 
\quad (U,V) \mapsto (\det(U_S)\times \det(V_S))_S\]
is Euclidean closed (and hence, Zariski closed).
A closely related map 
\[\F^{r \times n} \to \F^{n \choose r}, 
t\quad  U \mapsto (\det(U_S))_S\]
has been well-studied in algebraic geometry,
which gives the Pl\"{u}cker coordinates of elements in the Grassmannian variety. 
And hence, the image of this map is known to be a closed set. 
Putting it another way, our result says that
the set obtained by taking coordinatewise products of pairs of points in the Grassmannian variety is closed. 

Note that this is not a general phenomenon. 
It is easy to construct varieties where
the set obtained by taking coordinatewise products of pairs of points from the variety
is not closed. 
To see a simple example, consider the projective variety in $\mathbb{P}^2$ defined by
\[\{ [x : y : z] \mid  xz + y^2 - x^2 =0 \}.\]
Now, observe that the point $(0,1,0)$ cannot be obtained as a coordinatewise product of two points in the variety. On the other hand, it can be obtained as a limit of the product of two points $(\epsilon, 1, \epsilon-1/\epsilon)$  and $(1,1,0)$. 
See~\cite{BC22} for a related notion called
Hadamard power of varieties.

\paragraph*{Closure of a principal minor map.} Our main result also implies the closure of the image of a principal minor map, as defined below. 
The \textit{ affine principal minor map} 
$\phi: \C^{n^2}\xrightarrow{} \C^{2^n}$ is defined as 
\[\phi(A) = (\det(A_I))_{I \subseteq [n]}\]
where is $A_I$ is the principal submatrix of $A$ with rows and columns indexed by $I$. 
%
%For a matrix $A\in \C^{n\times n}$, let $A_*$ denote a vector in $\C^{2^n}$ with each coordinate indexed by a subset $I\subseteq [n]$. The coordinate indexed by set $I$ is equal to the minor of $A$ whose rows and columns are indexed by $I$. 
Lin and Sturmfels \cite{LIN20094121} showed that for any $n>0$, the image of $\phi$ 
%this affine principal minor map
on $n\times n$ matrices is closed. 
Our result implies the closure result for a  closely related map, which we refer to as the size $k$ principal minor map.
For any $k\leq n$, let us define the map
$\phi_k: \C^{n^2}\xrightarrow{} \C^{\binom{n}{k}}$ as
\[\phi_k(A) = (\det(A_I))_{I \in \binom{[n]}{k}}\]
where $\binom{[n]}{k}$ is the set of all size-$k$ subsets of $[n]$.
%For a matrix $A\in \C^{n\times n}$ with rank at most $k\leq n$, we consider the vector $A_*^k\in \C^ {\binom{n}{k}}$. Each  coordinate of $A_*^k$ corresponds to some subset $I\subset [n]$ of size $k$ and is equal to the minor of $A$ whose rows and columns are indexed by $I$. This lets us define a map  where $A\text{ is mapped to } A_*^k$. We call it  size $k$ principal minor map for $n\times n$ matrices.
We show that the image of $\phi_k$ on $n\times n$ rank-$k$ matrices is closed. Formally,
\begin{corollary}\label{theorem:minorClosure}
For any $n>0$ and $k\leq n$, the image of the size $k$ principal minor map on $n\times n$ matrices with rank at most $k$ is closed in $\C^{\binom{n}{k}}$.
\end{corollary}
One can define another similar map, where a rank-at-most-$k$ matrix is mapped to the tuple of its size-at-most-$k$ principal minors. Note that the closure of the image of this map follows easily from 
the result of Lin and Sturmfels \cite{LIN20094121}.
However, to the best of our knowledge, Corollary~\ref{theorem:minorClosure} does not follow from their result.

\paragraph*{Proof idea of the main result.}
As we said, our goal is to show that  the image of the  map 
\[(U,V) \mapsto (\det(U_S)\times \det(V_S))_S\]
is closed under approximation. 
The idea is to start with any two given matrices $U,V \in \F(\epsilon)^{r \times n}$ and construct 
matrices $\widehat{U}, \widehat{V} \in \F^{r \times n}$ such that for each size-$r$ subset $S\subseteq [n]$, we have
\[ \lim_{\e \to 0} \left( \det(U_S) \det(V_S) \right) =  \det(\widehat{U}_S) \det(\widehat{V}_S). \]
Of course, we can hope to construct such matrices only when the limit exists for every $S$. 
Note that one cannot simply 
apply the limit operation on the matrix entries because the matrix $U$ and $V$ can have 
 rational functions in $\epsilon$ as entries.

We can view each term like $\det(U_S)$ as a Laurent series in $\e$. 
For any Laurent series $f$ in $\epsilon$, one can define $\val(f)$ as the minimum exponent of $\epsilon$
appearing in $f$. 
Clearly, $\lim_{\e \to 0} f$
exists if and only if $\val(f) \geq 0$.
So let us assume that $\val(\det(U_S) \det(V_S)) \geq 0$ for every $S$. 
In other words,
\[\min_S \{ \val(\det(U_S) \det(V_S)) \} = \min_S \{ \val(\det(U_S)) + \val( \det(V_S)) \}  =0.\] 
Observe that only those sets $S$ which achieve this minimum will give a nonzero term in the limit. 
It would have been convenient if   min operator was distributive over the sum, i.e.,
\[\min_S \{ \val(\det(U_S)) + \val( \det(V_S)) \} = \min_S \{ \val(\det(U_S))\} + \min_S\{ \val( \det(V_S)) \},\]
but
that is of course not true.
Amazingly, it turns out that in the case of $\val$ function, it is almost true. 
This comes from the fact that the $\val$ 
function satisfies a matroid like exchange property: for any two distinct $S,T\subseteq[n]$ of size $r$ and any $j\in T\setminus S$, there exists a $k\in S\setminus T$ such that \[\val(\det(U_S))+\val(\det(U_T))\geq \val(\det(U_{S-k+j}))+\val(\det(U_{T-j+k})).
\]
%(see preliminaries in Section~\ref{sec:prelim}).
%
Based on this property, Dress and Wenzel~\cite{Dress90} defined the so-called valuated matroids. 
More interestingly, Murota~\cite{Murota96} proved the valuated matroid splitting theorem, which  says that the min operator indeed distributes over the sum of two $\val$ functions, but with a ``correction'' term which is a linear function. 
To be more precise, there is a tuple $\vect z \in \Z^{n}$ such that 
\begin{align*}\min_S \{ \val(\det(U_S)) + \val( \det(V_S))\}=& \min_S \{ \val(\det(U_S)) + \sum_{i \in S} {\vect z}_i\} 
\\
&+ \min_S\{ \val( \det(V_S))- \sum_{i \in S} {\vect z}_i \}.
\end{align*}
The correction term is easy to handle because of linearity. Then basically, the problem breaks into two independent problems on $U$ and $V$. 
That is, given any two  matrices $U,V \in \F(\epsilon)^{r \times n}$, construct 
matrices $\widehat{U}, \widehat{V} \in \F^{r \times n}$ such that for each size-$r$ subset $S\subseteq [n]$, we have
\[ \lim_{\e \to 0}  \det(U_S)   =  \det(\widehat{U}_S)
\text{ and } 
 \lim_{\e \to 0}  \det(V_S)   =  \det(\widehat{V}_S). \]
 The problem now becomes tractable essentially because the image of the map 
  $U \mapsto (\det(U_S))_S$ is known to be closed. 
 
\paragraph*{Discussion.} 
As discussed earlier, showing that a class of polynomials is closed under approximation also implies that it is Zariski closed. 
That is, the class of polynomials must be characterized by a set of polynomial equations (in the coefficients of the polynomials in the class). 
It would be interesting to find the set of characterizing equations for the class of determinant of symbolic matrices under rank one restriction.
Another natural class of polynomials for which we can study the closure question is that of
symbolic determinant under rank 2 (or higher) restriction.

%In order to prove this theorem, we use some  well known results from matroid theory. In the next section, we mention the definitions and results on matroids that we use. Then, in the third section we prove the above theorem.

%\paragraph{Proof Idea}
\section{Preliminaries and Notations}
\label{sec:prelim}
We use $\N$ to denote the set of natural numbers, $\R$ to denote the set of real numbers, $\C$ to denote the set of complex numbers, $\Z$ to denote the set of integers, and $\F$ to denote field $\R$ or $\C$, respectively. For a field $\F$ and an indeterminate $\e$, $\F(\e)$ denotes the fractional field. For a positive integer $n,$ $[n]$ denotes the set $\{1,2,\dots n\}$. For a set $E$, $2^E$ denotes the family of all possible subsets of $E$. For a subset $S$ of $E$ and an element $a\in E$, $S-a$ and $S+a$ denote the set $S\setminus \{a\}$ and $S\cup\{a\}$, respectively. For any subset $S$ of $[n]$, $\vect 1_S\in \F^n$ denotes the characteristic vector of the subset $S$. For a set $E$ and a non-negative integer $r$, $\binom{E}{r}$ denotes the set family consisting of all subsets of $E$ of size $r$. 

Every element $f$ in the fractional field $\F(\e)$ is of the form $g/h$ where $g,h\in \F[\e]$ with $h\neq 0$. For a nonzero polynomial $p\in \F[\e]$, let $\mindeg(p)$ be the degree of the minimum degree term in $p$. The function $\val$ from  $\F(\e)$ to $\Z$ is defined as 
\[
\val(f):=\left\{ 
\begin{array}{ll}
\mindeg(g) - \mindeg(h) & \text{ if } f\neq 0\\
+\infty                 & \text{otherwise}
\end{array}\right.
\] 
\begin{proposition}

\label{prop:val-property}
The $\val$ function  satisfies the following properties.
\begin{itemize}
    \item For any $f,g\in \F(\e),\ \val(fg)=\val(f)+\val(g)$.
    \item For any $f,g\in \F(\e)$, $\val(f+g)\geq \min\{\val(f),\val(g)\}$.
    \item For any $g\in \F(\e)\setminus\{0\},\ \val(1/g)=-\val(g)$.
    \item For an $f\in\F(\e)$, $\lim_{\e \to 0} f$ exists if and only if $\val(f)\geq 0$. Furthermore, $\lim_{\e \to 0} f=0 $ if and only if $\val(f)>0$.
\end{itemize}
\end{proposition}
For a polynomial $P\in \F(\e)[X]$ where $X=\{x_1,x_2,\dots ,x_n\}$ is the set of variables, we say $\lim_{\e\to 0} P$ exists if coefficient wise limit exists for every monomial of $P$ at $\e=0$. In other words, for any coefficient $f\in \F(\e)$ of a monomial of $P$, $\val(f)\geq 0$.\\ 
For a matrix $U\in \F^{r\times n}$, $i\in[r]$ and $j\in[n]$, $U[i,j]$ denotes the entry at $i$th row and $j$th column of $U$. For a matrix $U\in\F^{r\times n}$ and a subset $S\subseteq [n]$, $U_S$ denotes the submatrix of $U$ with columns indexed by $S$. 

% In the following lemma, we describe Cramer's rule used to solve system of linear equations. It will be helpful for our work.
% \begin{lemma}[Cramer's rule]
% \label{lem:cramer's_rule}
% Consider a system of $n$ linear equations for $n$ unknowns over a field $\F$, represented in the form of matrix multiplication as follows: $$A\x=\vect b,$$ where $A$ is a non-singular $n\times n$ matrix over $\F$, the vector $\x=(x_1,x_2,\ldots, x_n)^T$ is the column vector of variables and $\vect b$ is a column vector in $\F^n$. Then, for all $i\in[n]$, $$x_i=\frac{\det(A_i)}{\det(A)},$$ where $A_i$ is the matrix formed by replacing the $i$th column of $A$ by the column vector $\vect b$.
% \end{lemma}
% \SGnote{Add a reference for Cramer's rule.}

Next, we describe the Cauchy-Binet formula, which is an identity for the determinant of the product of two rectangular matrices of transposed shape. 
\begin{lemma}[Cauchy-Binet formula, \cite{Zeng93}]
\label{lem:Cauchy-Binet-formula}
Let $n \geq r$ be two positive integers. Let $A$ and $B$ are two $r\times n$ and $n\times r$ matrices over $\F$, respectively. Then $$\det(AB) = \sum_{S\in\binom{[n]}{r}}\det(A_S) \cdot \det(B_S),$$ where $B_S$ denotes the submatrix of $B$ with rows indexed by $S$.
\end{lemma}

Now we describe the Grassmann-Pl\"{u}cker identity.
\begin{lemma}[Equation 1.3 \cite{Dress92}]
\label{lem:grassman-identity}
Let $n\in\N$. Let $\vect a_0,\vect a_1,\ldots,\vect a_n, \vect b_2,\vect b_3,\ldots,\vect b_n$ be $2n$ vectors in $\F^n$. For all $i\in\{0,1,\ldots,n\}$, let $U_i$ and $V_i$ be the matrices $(\vect a_0,\ldots, \vect a_{i-1},\vect a_{i+1},\ldots,\vect a_n)$ and $(\vect a_i,\vect b_2,\ldots,\vect b_n)$, respectively. Then, $$\sum_{i=0}^n\det(U_i)\cdot\det(V_i)=0.$$ 
\end{lemma}
\paragraph*{Matroids.} A matroid $M$ is a set family $\calI$ defined on a ground set $E$ such that $\calI$ satisfies the following two properties:
\begin{enumerate}
    \item \textit{Closure under subsets:} If $X\in \mathcal{I}$ and $Y\subset X$, then $Y\in \mathcal{I}$.
    \item \textit{Augmentation Property:} If $|X|>|Y|$ and $X,Y\in \mathcal{I}$, then there exists $x\in X\setminus Y$ such that $Y\cup \{x\}\in \mathcal{I}$.
\end{enumerate}
The set family $\calI$ is called the \emph{independent set family} $M$. The augmentation property ensures that all the maximal independent sets of $M$ have the same size. The collection $\calB$ of all the maximal independent sets is called the \emph{base family} of $M$. The base family $\calB$ satisfies the following property:  
\begin{center}
\textit{Base exchange property:} Let $B,B'\in\calB$. Then for all $a\in B\setminus B'$ there exists a $b\in B'\setminus B$ such that $B-a+b$ is in $\calB$. 
\end{center}
Given the base family $\calB$ of a matroid $M$, its independent set family $\I=\{I\,\mid\,\exists B\in\calB,\  I\subseteq B\}$. Therefore, a matroid $M$ can be represented as $(E,\calI)$ or $(E,\calB)$. In this work, we mostly use $M=(E,\calB)$ to represent a matroid. Every matroid $M$ is associated with a function, $\rank:2^E\to \N$, defined as
\[
\rank(S)= \max \{|Y|\,\mid\, Y\subseteq S,\ Y\in \mathcal{I}\}.
\]
 The rank of the ground set $E$ is called the rank of the matroid $M$. It is equal to the cardinality of the bases. For more details on matroids, one can see some excellent textbooks like \cite{Oxl06, Sch03B}.

\paragraph*{Linear Matroids.} A well-known example of matroids is the \emph{linear matroids}. A linear matroid over a field $\F$ is represented by an $r\times n$ matrix $U$ over the field $\F$ with the full row rank. Assume that the columns are indexed by $[n]$, which is the ground set of the matroid. Let $\calB=\{B\subseteq [n]\,\mid\,|B|=r,\ \det(U_B)\neq 0\}$. It is not hard to prove that $M=([n],\calB)$ is a matroid with $\calB$ as the base family.

\paragraph*{Matroid Intersection.} Let $M_1=(E,\calB_1)$ and $M_2=(E,\calB_2)$ be two matroids defined on the same ground set $E$. The problem of finding a common base is called matroid intersection problem. The problem of  perfect matching for bipartite graphs and many other problems can be formulated in the language of the matroid intersection problem.
% \\ The matroid intersection theorem firstly given by Edmonds \cite{Edm70} states that the maximum size of a set in $\mathcal{I}_1\cap \mathcal{I}_2$ is equal to
% \[ \label{MIthm}
% \min \limits_{S\subseteq E}( rank_{M_1}(S)+rank_{M_2}(E\setminus S)).
% \]

In this paper, we study  symbolic matrix $M=\sum_{i=1}^{n}A_ix_i$ with each $A_i$  having  rank one. Next, we give an alternate representation of such symbolic matrices.
\begin{observation}\label{obs:CovertSymbolic} 
Let $M=\sum_{i=1}^{n}A_ix_i$ where each $A_i$ is a $r\times r$ rank one matrix over $\F$. Then, there exist $U,V\in \F^{r\times n}$ such that $M=UXV^T$ where $X$ is the $n\times n$ diagonal matrix with $x_i$ as its $i$th diagonal entry.
\end{observation}
\begin{proof}
Since $A_i$ is a rank one matrix over $\F$, there exist $\vect u^i,\vect v^i\in \F^r$ such that $A_i={\vect u^i}\cdot {\vect v^i}^T$. Let $U$ and $V$ be two $r\times n$ matrices such that the $i$th column of $U$ and $V$ are $\vect u^i$ and $\vect v^i$, respectively, for all $i\in [n]$. Then, for any $p,q\in[r]$, $UXV^T[p,q]=\sum_{i=1}^{n}u^i_pv^i_q x_i= \sum_{i=1}^{n}A_i[p,q]x_i$. This implies that $UXV^T=\sum_{i=1}^{n}A_ix_i$.
\end{proof}

\paragraph*{Valuated Matroid.} Dress and Wenzel \cite{Dress90, Dress92} introduced the notion of valuated matroid. Here, we discuss it as described by Murota \cite{Murota96}. Suppose that $M=(E, \calB)$ is a matroid with rank $r$. A \emph{valuation} on a matroid $M$ is a function $\omega$ from $\calB$ to $\infZ$ such that for all $B, B'\in \calB$ and $a\in B-B'$ there exists $b\in B'-B$ such that $B-a+b\in \calB$, $B'-b+a\in\calB$ and \begin{equation}
\label{eqn:val-Matroid}
\omega(B)+\omega(B')\geq \omega(B-a+b)+\omega(B'-b+a)    
\end{equation}
A matroid $M$ with a valuation function $\omega$ on it is called a \emph{valuated matroid}, and we denote it by the $3$-tuple $(E, \calB, \omega)$. The definition of the valuated matroids by Dress and Wenzel \cite{Dress90, Dress92} and in a subsequent work by Murota \cite{Murota96} consider the inequality in Equation \ref{eqn:val-Matroid} in the reverse direction. The reason is that their work talks about maximization problems over valuated matroids, but for our convenience, we describe their results in terms of minimization. %For a valuated matroid $M=(E,\calB,\omega)$ and a function $\vect w:E\to \Z$, $\omega[\vect w]$ is function from $\calB$ to $\infZ$, defined as follows: for all $B\in \calB$, $$\omega[\vect w](B)=\omega(B)+\sum_{b\in B}\vect w(b).$$

For a matrix, $U\in\F(\e)^{r\times n}$, the following lemma defines a valuation on the linear matroid represented by $U$. A very similar valuation has already been studied by Dress and Wenzel \cite{Dress92} and by Murota \cite[Example 3.2]{Murota96}. For an $f\in \F(\e)$ with $g,h\in \F[\e]$ and $f=g/h$, they consider $\deg_{\e}(f)$ instead of $\val(f)$ which is defined as the difference of degree of $p$ and $q$.
\begin{lemma}
\label{lem:val-to-valuation}
Let $U$ be an $r\times n$ matrix in $\F(\e)^{r\times n}$ with full row rank. Let $\calB$ be the base family of the linear matroid representable by $U$. Let $\omega$ be a function from $\calB$ to $\infZ$, defined as follows: for all $B\in \calB$, $$\omega(B)=\val(\det(U_B)).$$ Then $([n],\calB, \omega)$ forms a valuated matroid.
\end{lemma}
\begin{proof}
We prove the above lemma using the Grassmann-Pl\"{u}cker identity based technique used in \cite{Dress92}. From Grassmann-Pl\"{u}cker identity (Lemma \ref{lem:grassman-identity}), for any two distinct $S,T\subseteq[n]$ of size $r$ and any $j\in T\setminus S$, $$\det(U_S)\cdot \det(U_T)=\sum_{i\in S\setminus T}\mu_{i,j}\det(U_{S-i+j})\cdot \det(U_{T-j+i}),$$ where $\mu_{i,j}\in\{1,-1\}$. Then, from Proposition \ref{prop:val-property}, there exists a $k\in S\setminus T$ such that $$\val(\det(U_S))+\val(\det(U_T))\geq \val(\det(U_{S-k+j}))+\val(\det(U_{T-j+k})).$$
This implies that if $S,T\in\calB$, then for any $j\in T\setminus S$ there exists $k\in S\setminus T$ such that both $T-j+k$ and $S-k+j$ are in $\calB$ and $$\omega(S)+\omega(T)\geq \omega(S-k+j)+\omega(T-j+k).$$ Therefore, $([n],\calB, \omega)$ forms a valuated matroid.
\end{proof}
% \SGnote{The proof of the above lemma follows from the Grassman-Pl\"{u}cker identity. \SGnote{Add reference.} For a proof of the above lemma, one can see \cite{Dress92}. However, for their proof, we need to consider slightly different $\omega$. Instead of $\val(\cdot)$, $\omega(B)=\deg(\det(U_B))$, where $\deg(f/g)=\deg(f)-\deg(g)$ for two nonzero polynomials $f,g\in\F[\e]$, and $\deg(0)=-\infty$. However, their proof strategy also gives a proof of the above lemma in a straightforward way.}

Suppose that $U_1=(E, \calB_1,\omega_1)$ and $U_2=(E,\calB_2,\omega_2)$ are two valuated matroids over the same ground set $E$. Let $\vect w:E\to \Z$ be a weight function. For any weight function, $\vect w$ on the ground set $E$, it naturally extends to all the subsets of $E$ as follows: for any $S\subseteq E$, $\vect w(S)=\sum_{a\in S}\vect w(a).$ Then, the \emph{valuated matroid intersection} problem asks to find a common base $B\in\calB_1\cap\calB_2$ that minimizes $\vect w(B)+\omega_1(B)+\omega_2(B)$. Like Frank's weight splitting theorem for weighted matroid intersection \cite{Frank81}, Murota \cite[Theorem 4.2]{Murota96} gave a weight splitting theorem for the valuated matroid intersection.
% Wherever their result considers the maximization version of valuated matroid intersection, we describe it in terms of minimization, and the proof of this minimization version reduces to the maximization version in a natural way.
Here, we describe the result on the minimization version of valuated matroid intersection whose proof can be deduced from the result on the maximization version in a natural way. 
\begin{lemma}[Weight-splitting]
\label{lem:valuated-matroid-optimization}
Let $U_1=(E,\calB_1,\omega_1)$ and $U_2=(E,\calB_2,\omega_2)$ be two valuated matroids and $\vect w$ be a function from $E$ to $\Z$. Then, there exist $\vect w^1,\vect w^2:E\to \Z$ such that a common base $B$ minimizes $\vect w(B)+\omega_1(B)+\omega_2(B)$ if and only if the following holds:
\begin{enumerate}
\item $\vect w(e)=\vect w^1(e)+\vect w^2(e)$ for all $e\in E$.
\item $B$ is a minimum weight base for the matroid $U_1=(E,\calB_1)$ with respect to $\omega_1+\vect w^1$.
\item $B$ is a minimum weight base for the matroid $U_2=(E,\calB_2)$ with respect to $\omega_2+\vect w^2$.
\end{enumerate}
\end{lemma}
% \subsection{Arithmetic Circuits}
%  An Arithmetic circuit  over a field $\F$ and a set of variables $X=\{x_1,x_2,\dots x_n\}$ is a directed acyclic graph with leaf nodes labelled as a variable or an element from $F$. They are called input gates. The remaining nodes are either labelled as $+$ or $\times$ gates. A gate with 0 outdegree is called an output gate. It computes a polynomial in a natural way. For a leaf node labelled as $\alpha$, it computes the polynomial $\alpha$. For any other node, the polynomial computed by it is the sum or the product of the polynomials computed by its children depending on whether the node is labelled $+$ or $\times$. The number of gates is called the \textit{size} of the circuit and the maximum length of a path from an input to an output node exists as the \textit{depth} of the circuit. 
\section{Proof of our closure results}
%Theorems \ref{theorem:main} and \ref{theorem:minorClosure}}

In this section, we prove Theorem \ref{theorem:main} and Corollary \ref{theorem:minorClosure}. First, we discuss some lemmas that we use in the proof of our results.
One of the ingredients of our proof is the fact that the maximal minors of $r \times n$ matrices parameterize a variety (Pl\"{u}cker embedding of the Grassmannian). Since a variety is Euclidean closed, we get that 
for any $r\times n$ matrix $U$ over $\F(\e)$ whose $r \times r$ minors approach a vector $\vect u \in \F^{n \choose r}$ as $\e \rightarrow 0$,  there exists  an $r\times n$ matrix $\widehat U$ over $\F$ 
whose $r \times r$ minors equal to $\vect u$.
The next lemma shows how such a matrix $\widehat{U}$ can be constructed.  
%such that the limit value of the determinant of every $r\times r$ submatrix of $U$ at $\e = 0$ (if they exist) matches with the determinant of the respective submatrix of $\widehat{U}$. 
For notations, see Section \ref{sec:prelim}.

\begin{lemma}
\label{lem:remove-epsilon-one-matroid}
Let $U$ be a matrix in $\F(\epsilon)^{r\times n}$ such that for every $S\subseteq [n]$ of size $r$, $\lim_{\e\to 0} \det(U_S)$ exists. Then, we can construct $\widehat{U}$ in $\F^{r\times n}$ such that for every $S\subseteq [n]$ of size $r$ the following holds:
\[
\lim_{\e\to 0}\det(U_S) = \det(\widehat{U}_S).
\]
\end{lemma}

\begin{proof}
First consider the trivial case when $\lim_{\e\to 0} \det(U_S)$ is zero for every $S\subseteq [n]$ of size $r$. In that case, $\widehat U$ can be defined as the matrix with all entries being zero. Now, we assume that there exists a $S\subseteq[n]$ of size $r$ such that $\lim_{\e\to 0}\det(U_S)$ is non-zero. Without loss of generality, assume that  $\lim_{\e\to 0} \det(U_{[r]})$ is nonzero. Let $$U' = U^{-1}_{[r]}\cdot U.$$

\begin{claim}
\label{clm:validity-U'_S}
For every $S\subseteq [n]$ of size $r$, $\lim_{\e\to 0} \det(U'_S)$ exists.
\end{claim}
\begin{proof}
Since $U' = U^{-1}_{[r]}\cdot U$, for any $S\subseteq [n]$ of size $r$, $$\det(U'_S) = \det(U^{-1}_{[r]})\cdot \det(U_S).$$ Since $\det(U^{-1}_{[r]})=1/\det(U_{[r]})$ and  $\val(\det(U_{[r]})) = 0$, from Proposition \ref{prop:val-property}, $\val(\det(U^{-1}_{[r]}))$ is also zero. Therefore, applying Proposition \ref{prop:val-property}, we get that $\lim_{\e\to 0}\det(U^{-1}_{[r]})$ is non-zero. The hypothesis of the lemma ensures that $\lim_{\e\to 0}\det(U_S)$ exists.  Therefore, $$\lim_{\e\to0} \det(U'_S)=\lim_{\e\to 0}\det(U^{-1}_{[r]})\cdot \lim_{\e\to 0}\det(U_S).$$ This implies that $\lim_{\e\to 0}\det(U'_S)$ exists.
\end{proof}

\begin{claim}
\label{clm:val-of-U'}
For every $i\in [r]$ and $j\in [n]$, $\lim_{\e\to 0}U'[i,j]$ exists.
\end{claim}
\begin{proof}
% Let $A=U_{[r]}$. Then, applying the definition of $U'$ and Lemma \ref{lem:cramer's_rule}, for all $i\in[r]$ and $j\in[n]$, $$U'[i,j]=\frac{\det(A_{i})}{\det(A)},$$ where $A_{i}$ is the matrix formed by replacing the $i$th column of $U_{[r]}$ by the $j$th column of $U$. If $U'[i,j]=0$, then $\lim_{\e\to 0}U'[i,j]$ is trivially defined. Now, consider $U'[i,j]\neq 0$. This implies that both $\det(A_{i})$ and $\det(A)$ are nonzero. Therefore, from Proposition \ref{prop:val-property}, $$\val(U'[i,j])=\val(\det(A_i))-\val(\det(A)).$$ Since $A_i=U_{T}$ where $T=[r]-\{i\}+\{j\}$, the hypothesis of the lemma ensures that $\lim_{\e\to 0}\det(A_i)$ exists. Hence, from Proposition \ref{prop:val-property}, $\val(\det(A_i))\geq 0$. On the other hand, applying Proposition \ref{prop:val-property}, $\val(\det(A))=0$ since $\lim_{\e\to 0}\det(U_{[r]})$ is nonzero. Therefore, $\val(U'[i,j])\geq 0$. This combined with Proposition \ref{prop:val-property} imply that $\lim_{\e\to 0}\val(U'[i,j])$ exists.  

From the definition, $U' = [I_r | A ]$ where $I_r$ is the $r\times r$ identity matrix. The claim trivially follows for $i,j\in [r]$. For an $i\in [r]$ and $j\in [n] - [r]$, let $T = [r] - \{i\} + \{j\}$, and $U'_T$ be the matrix obtained by replacing the $i$th column of $I_r$ by the $j$th column of $U'$. This implies that the matrix $U'_T$ is of the following form:
\[
\quad
U'_T = 
\begin{bmatrix}
1 &0 &0 &\ldots &U'[1,j] &\ldots &0\\
0 &1 &0 &\ldots &U'[2,j] &\ldots &0\\
\vdots  &\vdots &\vdots &\ddots &\vdots &\vdots &\vdots\\
0 &0 &0 &\ldots &U'[i,j] &\ldots &0\\
\vdots &\vdots &\vdots &\vdots &\vdots &\ddots &\vdots\\
0 &0 &0 &\ldots &U'[r,j] &\ldots &1
\end{bmatrix}.
\]
Therefore, $\det(U'_T) = U'[i,j]$. From the hypothesis of the lemma combined with Proposition \ref{prop:val-property}, we know that $\val(\det(U'_T)) \geq 0$. Hence, $\val(U'[i,j]) \geq 0$. Now applying Proposition \ref{prop:val-property}, $\lim_{\e\to 0}U'[i,j]$ exists.
\end{proof}

Now we define the matrix $\widetilde U\in \F^{r\times n}$ as follows: for all $i\in[r]$ and $j\in[n]$, $$\widetilde U[i,j] := \lim_{\e\to 0}U'[i,j].$$ From Claim \ref{clm:val-of-U'}, the entries of the matrix $\widetilde U$ are well defined. Since determinant is a continuous function, 
\begin{equation}\label{eqn:lemma-2_1}
\lim_{\e\to 0}\det(U'_S)=\det(\widetilde U_S).
\end{equation}
% Next we show that for any $S\subseteq[n]$ of size $r$, 
% \begin{equation}
% \label{eqn:lemma-2_1}
% \lim_{\e\to 0}\det(U'_S)=\det(\widetilde U_S).    
% \end{equation} 
% Consider a relabeling of the elements of $S$ by the elements of $[r]$. Then, $$\lim_{\e\to 0}\det(U'_S) = \lim_{\e\to 0}\left(\sum_{\sigma\in \sym{r}}(-1)^{\mathrm{sgn}(\sigma)}\prod_{i\in[r]}U'_S[i,\sigma(i)]\right).$$
% Applying Claim \ref{clm:val-of-U'}, 
% $$\lim_{\e\to 0}\det(U'_S) = \sum_{\sigma\in \sym{r}}(-1)^{\mathrm{sgn}(\sigma)}\prod_{i\in[r]}\lim_{\e\to 0}U'_S[i,\sigma(i)].$$ From the definition of $\widetilde U$, 
% \begin{align*}
% \lim_{\e\to 0}\det(U'_S) &= \sum_{\sigma \in\sym{r}}(-1)^{\mathrm{sgn}(\sigma)}\prod_{i\in[r]}\widetilde U[i,\sigma(i)]\\
% &= \det(\widetilde U_S).
% \end{align*}

Let $\lim_{\epsilon\to 0} det (U_{[r]}) = \alpha$. Consider the matrix $\widehat{U}\in \F^{r\times n}$ which exists by multiplying the first row of $\widetilde U$ by $\alpha$, that is for all $i\in[r]$ and $j\in[n]$,
\[
\widehat{U}[i,j] = \left\{
\begin{array}{ll}
\alpha\cdot \widetilde{U}[i,j]  & \textrm{ if }i=1\\
\widetilde{U}[i,j] & \textrm{otherwise}.
\end{array}
\right.
\] 
The definition of $\widehat U$ implies that for any $S\subseteq [n]$ of size $r$,
\begin{equation}
\label{eqn:lemma-2_2}
\det(\widehat U_S)=\alpha\cdot \det(\widetilde U_S). 
\end{equation}
From the definition of $U'$, $$\lim_{\e \to 0} \det(U_S) = \lim_{\e\to 0}(\det(U_{[r]}\cdot \det(U'_S)).$$ Applying Claim \ref{clm:validity-U'_S}, $\lim_{\e\to 0}\det(U'_S)$ exists. Therefore,
\begin{align*}
\lim_{\e\to 0}\det(U_S) &= \lim_{\e\to 0}\det(U_{[r]})\cdot \lim_{\e\to 0}\det(U'_S)\\
&= \alpha\cdot \det(\widetilde U_S)\quad& \text{[from Equation \ref{eqn:lemma-2_1}]}\\
&= \det(\widehat U_S)\quad& \text{[from Equation \ref{eqn:lemma-2_2}]}.
\end{align*}
This completes the proof of our lemma.
\end{proof}

Suppose that $U,V$ are two matrices in $\F(\e)^{r\times n}$ with full row rank. Let $\lim_{\e \to 0}(\det(U_S)\cdot \det(V_S))$ exists for all $S\subseteq [n]$ of size $r$. However, the limit value of  $\det(U_S)$ and $\det(V_S)$ at $\e=0$ individually may not exist for all $S$. Our next lemma shows that there exists two $r\times n$ matrices $\widetilde U$ and $\widetilde V$ such that the limit value of both $\det(U_S)\cdot \det(V_S)$ and $\det(\widetilde U_S)\cdot \det(\widetilde V_S)$ at $\e=0$ are same and also the limit value of  $\det(\widetilde U_S)$ and $\det(\widetilde V_S)$ at $\e=0$ individually exists.

%For a matrix $U\in \F(\e)^{r\times n}$ and a vector $\vect{v}=(v_1,v_2,\dots,v_n)^T\in \F(\e)^{n}$, $\mvp{U}{\vect{v}}$ represents a new matrix where the $i$th column is the $i$th column of $U$ multiplied by $v_i$. In other words, for any $i\in[r]$ and $j\in[n]$, $(\mvp{U}{v})[i,j]=v_jU[i,j]$. For an $\vect{a}=(a_1,a_2,\ldots, a_n)^T \in \Z^n$, $\e^{\vect{a}}$ represents the vector $(\e^{a_1},\e^{a_2},\ldots,\e^{a_n})^T$.

For a matrix $U\in \F(\e)^{r\times n}$ with full row rank, let us define
%$$\mval(U) := \min \limits_{S\in\binom{[n]}{r}\,:\,\det(U_S)\neq 0} \val(det(U_S)).$$
$$\mval(U) := \min \limits_{S\in\binom{[n]}{r}} \val(\det(U_S)).$$

\begin{lemma}
\label{lem:removing-epsilon-MI}
Let $U,V$ in $\F(\e)^{r\times n}$ with full row rank. Let $\lim_{\e \to 0}\det(U_S)\cdot \det(V_S)$ exists for all $S\subseteq [n]$ of size $r$. Then, there exist $\widetilde{U}, \widetilde{V}$ in $\F(\epsilon)^{r\times n}$ such that for every $S\subseteq [n]$ of size $r$ the following holds:
\[
\lim_{\epsilon\to 0} \det(U_S)\det(V_S) =  \left( \lim_{\epsilon \to 0}\det(\widetilde{U}_S) \right)\cdot \left( \lim_{\epsilon \to 0}\det(\widetilde{V}_S) \right).
\]
\end{lemma}
\begin{proof}
When $\lim_{\e\to 0} \det(U_S)\det(V_S)=0$ for all $S\subseteq [n]$ of size $r$, the lemma is trivial to prove. Now we consider the case  when there exists an $S\subseteq [n]$ of size $r$ such that $\lim_{\e\to 0} \det(U_S)\det(V_S)\neq0$. Next, we show that there exists a vector $\vect z\in \Z^n$ such that 
$$\mval({U} \cdot \diag(\e^{\vect z}))+\mval({V} \cdot \diag(\e^{-\vect z}) )=0,$$
where $\diag(\e^{\vect z})$ is the diagonal matrix with $(i,i)$th entry as $\epsilon^{z_i}$.
Let $\calB_1$ and $\calB_2$ be the base families for the linear matroid represented by $U$ and $V$, respectively. Let $\omega_1$ be a function from $2^{[n]}$ to $\infZ$ defined as follows: for all $B\in2^{[n]}$,
\[
\omega_1(B)=\left\{
\begin{array}{ll}
\val(\det(U_B)) & \text{if } B\in\calB_1\\
+\infty         & \text{otherwise}
\end{array}
\right.
\]
Similarly, we can define $\omega_2:2^{[n]}\to\infZ$ for the matrix $V$. Now, from Lemma \ref{lem:val-to-valuation}, both $([n],\calB_1,\omega_1)$ and $([n],\calB_2,\omega_2)$ are valuated matroids. Therefore, applying Lemma \ref{lem:valuated-matroid-optimization} with $\vect w$ as the zero function on $[n]$, there exists a weight function $\vect z:[n]\xrightarrow{}\Z$  such that a common base $B\in\calB_1\cap\calB_2$ minimizes $\omega_1(B)+\omega_2(B)$ if and only if the following holds:
\begin{enumerate}
\item $B$ is a minimum weight base for the matroid $([n], \calB_1)$ with respect to  $\omega_1+\vect z$.
\item $B$ is a minimum weight base for the matroid $([n],\calB_2)$ with respect to $\omega_2-\vect z$.
\end{enumerate}
Abusing notation, let $\vect z$ also denote a vector in $ \Z^n$ with $i$th coordinate as $\vect z(i)$. Let $U'={U} \cdot \diag({\e^{\vect z}})$ and $V'={V} \cdot \diag({\e^{-\vect z}})$. From the definitions, $\mval(U')$ is the minimum weight of a base of $([n],\calB_1)$ with respect to $\omega_1+\vect z$. Similarly, $\mval(V')$ is the minimum weight of a base of $([n], \calB_2)$ with respect to $\omega_2-\vect z$. Since for every $S\subseteq [n]$ of size $r$, $\lim_{\e\to 0}\det(U_S)\det(V_S)$ exists, for all $B\in\calB_1\cap\calB_2$, $\val(\det(U_B))+\val(\det(V_B))\geq 0$. On the other hand, from our assumption, there exists an $S\subseteq[n]$ of size $r$ such that $\lim_{\e\to 0}\det(U_S)\det(V_S)\neq 0$. Therefore, $$\min_{B\in\calB_1\cap\calB_2}\val(\det(U_B))+\val(\det(V_B))=0.$$ This implies that 
\begin{align*}
\mval(U')+\mval(V') &=\min_{B\in\calB_1}(\omega_1+\vect z)(B)+\min_{B\in\calB_2}(\omega_2-\vect z)(B)\\
&=\min_{B\in\calB_1\cap\calB_2}\omega_1(B)+\omega_2(B)\\
&=\min_{B\in\calB_1\cap\calB_2}\val(\det(U_B))+\val(\det(V_B))\\
&=0.
\end{align*}
Let $c= \mval(U')= - \mval(V')$. Let $\widetilde{U}$ and $\widetilde{V}$ be the matrix obtained by multiplying  the first row of $U'$ and $V'$ by $\e^{-c}$ and $\e^c$, respectively. Thus, for all $S\subseteq [n]$ of size $r$, we have that $$\det(U_S)\cdot \det(V_S)=\det(U'_S)\cdot \det(V'_S)= \det (\widetilde{U}_S)\cdot \det(\widetilde{V}_S),$$ and $\mval(\widetilde{U})=\mval(U')-c=0$. Similarly, $\mval(\widetilde{V})=0$. This implies that for all $S\subseteq [n]$ of size $r$, $$\lim_{\e \to 0} \det(U_S)\cdot \det(V_S) = \left(\lim_{\e\to 0}\det(\widetilde{U}_S) \right)\cdot  \left(\lim_{\e \to 0}\det(\widetilde{V}_S) \right).$$
\end{proof}
\subsection{Proof of Theorem \ref{theorem:main}}
In this subsection, we give the proof of Theorem \ref{theorem:main}. First, we prove  for the case when $A_0 = 0$. From Observation~\ref{obs:CovertSymbolic}, we get $U,V \in \F(\epsilon)^{r\times n}$ such that $\sum_{i=1}^n A_ix_i = UXV^T$ where $X$ is the diagonal matrix with $x_i$ as its $i$th diagonal entry. Abusing notation, we use $X_S$ to denote $\prod_{i\in S}x_i$. Therefore,
\begin{align*}
f &= \lim_{\e\to 0}\det\left(\sum_{i=1}^n A_ix_i\right) = \lim_{\e\to 0}\det(UXV^T) \\
&= \lim_{\e\to 0}\sum_{S\subseteq [n], |S| = r} \det(U_S)\det(V_S)X_S &\text{[from Lemma \ref{lem:Cauchy-Binet-formula}]}\\
&= \sum_{S\subseteq [n], |S| = r}\left(\lim_{\e\to 0}\det(U_S)\det(V_S)\right)X_S. 
\end{align*}
In the last equality above, we can take the limit inside as $f$ is defined if and only if the limit exists for  the coefficient of every monomial.
Applying Lemma \ref{lem:removing-epsilon-MI},
$$f= \sum_{S\subseteq [n],\, |S| = r}\left(\lim_{\e\to 0}\det(\widetilde{U}_S)\right)\left(\lim_{\e\to 0}\det(\widetilde{V}_S)\right)X_S.$$
From Lemma \ref{lem:remove-epsilon-one-matroid}, we have two $r\times n$ matrices $\widehat U$ and $\widehat V$ in $\F^{r\times n}$ such that
\begin{align*}
f &= \sum_{S\subseteq [n],\, |S| = r}\det(\widehat{U}_S)\det(\widehat{V}_S)X_S\\
&=\det(\widehat{U}X\widehat{V}^T).
\end{align*}
For all $i\in[n]$, let $B_i$ be the $r\times r$ rank one matrix defined as $\widehat{U}[i]\cdot \widehat{V}[i]^T$, where $\widehat{U}[i]$ and $\widehat{V}[i]$ are the $i$th columns of $\widehat U$ and $\widehat V$ respectively. Then, $$f= \det(\widehat{U}X\widehat{V}^T) = \det(\sum_{i=1}^{n}B_ix_i).$$  This completes the proof of Theorem \ref{theorem:main} where $A_0 = 0$.

For the case when $A_0\neq 0$, we first give the following lemma that essentially reduces it to the previous case. The proof idea comes from Anderson, Shpilka, and Volk~\cite{PC} (as cited in \cite[Lemma~4.3]{GT20}).

For positive integers $m$ and $n$, let $I_n$ denote the $n\times n$ identity matrix and $0_{m,n}$ denote the $m\times n$ rectangular matrix with all zeros.
\begin{lemma}\label{lemma:constant-part}
Let $P = \det\left( A_0 + UXV^T \right)$ for some $U,V$ in $\F(\e)^{r\times n}$, $A_0\in \F(\e)^{r\times r}$ and $X$ is an $n\times n$ diagonal matrix with $x_1, x_2, \ldots, x_n$ in the diagonal. Let $X'$ be a $(2n+r)\times (2n+r)$ diagonal matrix with $x_1,x_2,\dots x_{2n+r}$ in the diagonal. Then, there exist rectangular matrices $U', V' \in $ $\F(\e)^{(n+r)\times (2n+r)}$  such that the following holds:
\begin{itemize}
    \item Let $Q$ be the polynomial in $x_1,x_2,\ldots x_n$ obtained by putting $x_{n+1},\ldots, x_{2n+r}$ equal to 1 in $det(U'X'V'^T)$. Then, $P=Q$.
    \item  If $\lim_{\e \to 0}P $ exists, then $\lim_{\e \to 0}\det(U'X'V'^T)$ also exists.
\end{itemize} 
\end{lemma}
\begin{proof}
Let us define
\[
U' = \left[
\begin{array}{c c c | c c c| c}
&&&&&&\\
&0_{n,n}&&&I_n&&V^T\\
&&&&&&\\
\hline
&-U&&&0_{r,n}&&A_0
\end{array}
\right]
\quad\quad\text{and, }
V' = \left[\begin{array}{c c c | c c c| c}
&&&&&&\\
&I_n&&&I_n&&0_{n,r}\\
&&&&&&\\
\hline
&0_{r,n}&&&0_{r,n}&&I_r
\end{array}
\right].
\]
Let $X_1$ be a $n\times n$ diagonal matrix with $x_{n+1},\ldots,x_{2n}$ in the diagonal and $X_2$ be a $r\times r$ diagonal matrix with $x_{2n+1},\ldots, x_{2n+r}$ in the diagonal.
We now consider $U'X'V'^T$. Notice that,
\[
U'X'V'^T =
\left[
\begin{array}{c c c | c c c| c}
&&&&&&\\
&0_{n,n}&&&X_1&&V^TX_2\\
&&&&&&\\
\hline
&-UX&&&0_{r,n}&&A_0X_2
\end{array}
\right]
\cdot \left[\begin{array}{c c c | c c c}
&&&&\\
&I_n&&&0_{n,r}\\
&&&&\\ 
\hline
&&&&\\
&I_n&&&0_{n,r}\\
&&&&\\
\hline
&0_{r,n}&&&I_r
\end{array}
\right] = \begin{bmatrix}
X_1 &V^TX_2\\
-UX &A_0X_2
\end{bmatrix}. 
\]
Let $A,B,C,D$ be matrices where $A$ and $D$ are square matrices and $A$ is invertible. Then, we have
\[\det\left(\begin{array}{cc}
    A & B \\
    C & D
\end{array} \right) = \det(A).\det(D-CA^{-1}B)\] 
 Therefore, 
 \begin{align*}\det(U'X'V'^T) &= \det(X_1). \det(A_0X_2 + UXX_1^{-1}V^TX_2) \\
 &= \det(X_1).\det(A_0 + UXX_1^{-1}V^T).\det(X_2).
 \end{align*}
 It is easy to see that if we put the value of 1 to $x_{n+1},\ldots,x_{2n+r}$, we get $\det(A_0+UXV^T)$. Also,
 \[ \lim_{\e \to 0}\det(U'X'V'^T)=\det(X_1).\det(X_2).\lim_{\e \to 0} \det(A_0 + UXX_1^{-1}V^T).\]
The second part of the lemma follows from the fact that if $\lim_{\e \to 0}P $ exists, then $\lim_{\e \to 0} \det(A_0 + UXX_1^{-1}V^T)$ also exists as $XX_1^{-1}$ can be treated as a diagonal matrix with a different set of indeterminates.
\end{proof}
Now we prove for the case when $A_0\neq 0$. Let $f=\det(A_0+UXV^T)$ and $f'= \det (U'X'V'^T)$. From the Lemma \ref{lemma:constant-part}, $\lim_{\e\to 0}f'$ exists as it is given that $\lim_{\e \to 0}f$ exists. Just like we discussed above for the case of $A_0=0$, we can get $\widehat{U}',\widehat{V}'\in \F^{(n+r)\times (2n+r) }$ such that $\lim_{\e\to 0} f'= \det(\widehat{U}'X'\widehat{V}'^T)$. For all $i\in[2n+r]$, let $B_i$ be the $(n+r)\times (n+r)$ rank one matrix defined as $\widehat{U}'[i]\cdot \widehat{V}'[i]^T$, where $\widehat{U}'[i]$ and $\widehat{V}'[i]$ are the $i$th columns of $\widehat {U}'$ and $\widehat {V}'$ respectively. Hence, $\lim_{\e\to 0}f'=\det(\sum_{i=1}^{2n+r}B_ix_i)$. Let $\sum_{i=n+1}^{2n+r}B_i = B_0$. From the first part of Lemma \ref{lemma:constant-part}, $\lim_{\e\to 0}f= \det(B_0+\sum_{i=1}^{n}B_ix_i)$.

\subsection{Proof of Corollary \ref{theorem:minorClosure}}
 We will show the following lemma which directly implies Corollary~\ref{theorem:minorClosure}. 
 \begin{lemma}
     Let $A\in\C(\e)^{n\times n}$ be a matrix of rank at most $k$ and $A[S]$ denote the minor of $A$ whose rows and columns are indexed by $S\subseteq [n]$. Let $\lim_{\e \to 0} A[S]$ exist for all subset $S\subset [n]$ of size $k$. Then, there exists $B\in \C^{n\times n}$ such that for all $S\subset [n]$ of size $k$,
     \[\lim_{\e\to 0} A[S]=B[S]\]
 \end{lemma}
 \begin{proof}
     The claim is trivial when $\rank(A)<k$ as all the minors are zero. Hence, we assume that $\rank(A)=k$. Let $U,V\in \C(\e)^{k\times n }$ such that $U^T,V$ is a rank-factorization of $A$. This implies that $A=U^T.V$ and for any subset $S\subset [n]$, $A[S]=\det(U_S^T.V_S)$. Since $\lim_{\e \to 0}A[S]=\lim_{\e \to 0} \det(U_S)\det(V_S)$ exists for all $S\subset [n]$ of size $k$, from Lemma \ref{lem:removing-epsilon-MI} there exists $\widetilde{U}, \widetilde{V}$ in $\C(\epsilon)^{r\times n}$ such that for every $S\subseteq [n]$ of size $k$ the following holds:
\[
\lim_{\epsilon\to 0} A[S] =  \left( \lim_{\epsilon \to 0}\det(\widetilde{U}_S) \right)\cdot \left( \lim_{\epsilon \to 0}\det(\widetilde{V}_S) \right).
\]
From Lemma \ref{lem:remove-epsilon-one-matroid}, there exist two $k\times n$ matrices $\widehat{U}$ and $\widehat{V}\in \C^{k\times n}$ such that for all $S\subset [n]$, the following holds:
\[\lim_{\e \to 0} \det(\widetilde{U}_S)= \det(\widehat{U}_S) \text{ and }  \lim_{\e \to 0} \det(\widetilde{V}_S)= \det(\widehat{V}_S) \]
Let $B=\widehat{U}^T.\widehat{V}$. Hence, for all $S\subset [n]$ of size $k$,
\[\lim_{\e\to 0}A[S]= \det(\widehat{U}_S^T).\det(\widehat{V}_S)= \det(\widehat{U}_S^T.\widehat{V}_S)=B[S].\]
 \end{proof}
\bibliographystyle{alpha}
\bibliography{borderVDet1}
% \include{MimickingNetworks}
% \bibliographystyle{alpha}
% \bibliography{matroidIntersection,MimickingNetworksReferences}
\end{document}